\documentclass[10pt,a4paper]{article}
\usepackage{amsfonts,amsmath,amssymb,cite}
\newcommand{\Real}{\mathop{\textrm{Re}}}

\begin{document}
\title{Two-dimensional hydrogen-like atom \\ 
in a weak magnetic field}
\author{Rados{\l}aw Szmytkowski\footnote{Email:
radoslaw.szmytkowski@pg.edu.pl} \\*[3ex]
Atomic and Optical Physics Division, \\
Department of Atomic, Molecular and Optical Physics, \\
Faculty of Applied Physics and Mathematics,
Gda{\'n}sk University of Technology, \\
ul.\ Gabriela Narutowicza 11/12, 80--233 Gda{\'n}sk, Poland}
\date{}
\maketitle
\begin{abstract} 
We consider a non-relativistic two-dimensional (2D) hydrogen-like
atom in a weak, static, uniform magnetic field perpendicular to the
atomic plane. Within the framework of the Rayleigh--Schr{\"o}dinger
perturbation theory, using the Sturmian expansion of the generalized
radial Coulomb Green function, we derive explicit analytical
expressions for corrections to an arbitrary planar hydrogenic
bound-state energy level, up to the fourth order in the strength of
the perturbing magnetic field. In the case of the ground state, we
correct an expression for the fourth-order correction to energy
available in the literature.
\mbox{} \\*[3ex]
\mbox{} \\*[1ex]
\noindent
\textbf{PACS 2010:} 03.65.Ge, 31.15.xp, 32.60.+i, 71.35.Ji
\end{abstract}
%
%
\section{Introduction}
\label{I}
\setcounter{equation}{0}
Theoretical studies on elementary two-dimensional quantum structures
in magnetic fields have been carried out for several decades
\cite{Akim67,Gork68,Shin70,MacD86,Adam88,Edel89,Zhu90a,Zhu90b,Mart92,%
Le93,Lehm95,Taut95,Vill98a,Vill98b,Ho00,Vill01,Robn03,Vill03,Soyl06,%
Soyl07}. Recent spectacular developments in single-layer materials
science have given a fresh impact to such investigations
\cite{Rutk09,Posz10,Gade11,Posz11,Hoan13a,Hoan13b,Posz14,Esco14a,%
Esco15,Esco14b,Fera15,Flav15,Liu15,Arde16,Hoan16,Le17}. In result of
the work done so far, at the present moment we understand some
aspects of magnetic-field-induced properties of 2D analogues of atoms
and molecules, but our knowledge on the subject still appears to be
far from being complete.

Recently, we have come across a need to know exact analytical
representations for low-order perturbation-theory corrections to an
arbitrary energy level of a two-dimensional analogue of a
hydrogen-like atom placed in a weak and uniform magnetic field
perpendicular to the atomic plane. The first-order correction may be
obtained trivially for any atomic state. Exact values of the
second-order corrections for states with the principal quantum
numbers $1\leqslant n\leqslant4$ may be derived from a table provided
in Ref.\ \cite{MacD86}. The third-order correction may be shown to
vanish identically for any state (in fact, the same happens for all
odd-order corrections other than the first-order one), while in
Refs.\ \cite{Hoan13a,Fera15} the fourth-order correction has been
given, but for the ground level only. Approximate expressions for
several higher even-order corrections to states with zero radial
quantum numbers and with principal quantum numbers not exceeding six
are contained in Ref.\ \cite{Hoan16}. However, neither of the
publications invoked above, nor any other related one we have had in
hands in the course of browsing the literature, contains the general
formulas we have been seeking for. This is a bit astonishing in view
of the fact that for a similar problem of the planar one-electron
atom placed in a weak, uniform, in-plane electric field, closed-form
analytical expressions for Stark--Lo Surdo corrections to energies of
discrete parabolic eigenstates are known up to the sixth order in the
perturbing field \cite{Fern92,Adam92}. Under the circumstances, we
have derived expressions for the second- and fourth-order
magnetic-field-induced corrections to an arbitrary energy level of
the planar hydrogenic atom. The results of that study are presented
in this work. We believe they may be of some interest, in particular
because the result for the fourth-order correction to the ground
state given in Ref.\ \cite{Hoan13a}, and then repeated in Ref.\
\cite{Fera15}, has been found to be incorrect.
%
%
\section{Preliminaries}
\label{II}
\setcounter{equation}{0}
We consider a one-electron atom with a point-like and spinless
nucleus at rest. The electron is constrained to move in a plane
through the nucleus. A potential of interaction between the nucleus
(of charge $+Ze$) and electron (of charge $-e$ and mass $m$) is taken
to be the attractive Coulomb one-over-distance one. The system is
subjected to the action of a static uniform magnetic field of
induction $\boldsymbol{B}$, which is perpendicular to the atomic
plane. With the electron radius vector $\boldsymbol{r}$ being
referred to the nucleus, the two-dimensional time-independent
Schr{\"o}dinger equation for the electron is
\begin{equation}
\left\{\frac{[-\mathrm{i}\hbar\boldsymbol{\nabla}
+e\boldsymbol{A}(\boldsymbol{r})]^{2}}{2m}
-\frac{Ze^{2}}{(4\pi\epsilon_{0})r}\right\}\Psi(\boldsymbol{r})
=E\Psi(\boldsymbol{r})
\qquad (\boldsymbol{r}\in\mathbb{R}^{2}),
\label{2.1}
\end{equation}
where $r=|\boldsymbol{r}|$ and $\boldsymbol{A}(\boldsymbol{r})$ is a
vector potential of the magnetic field. Equation (\ref{2.1}) is to be
solved, with the electron energy $E$ chosen as an eigenvalue, subject
to the constraint that the wave function $\Psi(\boldsymbol{r})$ is
single-valued and bounded for all $\boldsymbol{r}\in\mathbb{R}^{2}$,
including the point $\boldsymbol{r}=0$ and the point at infinity.

Throughout this paper, we shall be working in the symmetric gauge, in
which the vector potential $\boldsymbol{A}(\boldsymbol{r})$ is
\begin{equation}
\boldsymbol{A}(\boldsymbol{r})
=\frac{1}{2}\boldsymbol{B}\times\boldsymbol{r}.
\label{2.2}
\end{equation}
Then, the Schr{\"o}dinger equation (\ref{2.1}) may be rewritten as
\begin{equation}
\left[-\frac{\hbar^{2}}{2m}\boldsymbol{\nabla}^{2}
+\frac{e\hbar}{2m}\boldsymbol{B}\cdot\boldsymbol{\Lambda}
+\frac{e^{2}B^{2}r^{2}}{8m}
-\frac{Ze^{2}}{(4\pi\epsilon_{0})r}\right]\Psi(\boldsymbol{r})
=E\Psi(\boldsymbol{r}),
\label{2.3}
\end{equation}
where
\begin{equation}
\boldsymbol{\Lambda}
=-\mathrm{i}\boldsymbol{r}\times\boldsymbol{\nabla}
\label{2.4}
\end{equation}
is a (dimensionless) orbital angular momentum operator for the
electron. The form of the Hamiltonian operator in the Schr{\"o}dinger
equation (\ref{2.3}) suggests one introduces the polar coordinates
$r$ and $\varphi$, with $0\leqslant r<\infty$ and
$0\leqslant\varphi<2\pi$; Eq.\ (\ref{2.3}) is then transformed into
the following one:
\begin{equation}
\left[-\frac{\hbar^{2}}{2m}
\left(\frac{\partial^{2}}{\partial r^{2}}+
\frac{1}{r}\frac{\partial}{\partial r}
+\frac{1}{r^{2}}\frac{\partial^{2}}{\partial\varphi^{2}}\right)
-\frac{\mathrm{i}e\hbar B}{2m}\frac{\partial}{\partial\varphi}
+\frac{e^{2}B^{2}r^{2}}{8m}
-\frac{Ze^{2}}{(4\pi\epsilon_{0})r}\right]\Psi(r,\varphi)
=E\Psi(r,\varphi).
\label{2.5}
\end{equation}
The benefit from the use of the polar coordinates is that Eq.\
(\ref{2.5}) is separable, in the sense that it possesses particular
solutions of the form
\begin{equation}
\Psi_{nlm_{l}}(r,\varphi)
=\frac{1}{\sqrt{r}}P_{nlm_{l}}(r)\Phi_{m_{l}}(\varphi)
\qquad (l=|m_{l}|),
\label{2.6}
\end{equation}
where
\begin{equation}
\Phi_{m_{l}}(\varphi)=\frac{1}{\sqrt{2\pi}}
\mathrm{e}^{\mathrm{i}m_{l}\varphi}
\qquad (m_{l}\in\mathbb{Z}).
\label{2.7}
\end{equation}
Plugging Eq.\ (\ref{2.6}) into Eq.\ (\ref{2.5}) and exploiting Eq.\
(\ref{2.7}) yields the radial Schr{\"o}dinger equation
\begin{subequations}
\begin{equation}
\left[-\frac{\hbar^{2}}{2m}\frac{\mathrm{d}^{2}}{\mathrm{d}r^{2}}
+\frac{\hbar^{2}\left(l^{2}-\frac{1}{4}\right)}{2mr^{2}}
+m_{l}\frac{e\hbar B}{2m}+\frac{e^{2}B^{2}r^{2}}{8m}
-\frac{Ze^{2}}{(4\pi\epsilon_{0})r}\right]P_{nlm_{l}}(r)
=E_{nlm_{l}}P_{nlm_{l}}(r),
\label{2.8a}
\end{equation}
which is to be solved subject to the boundary conditions
\begin{equation} 
\textrm{$P_{nlm_{l}}(r)/\sqrt{r}$ bounded
for $r\to0$ and for $r\to\infty$}. 
\label{2.8b} 
\end{equation}
It is easy to deduce from the standard asymptotic analysis that for
$B\neq0$ the constraints displayed in Eq.\ (\ref{2.8b}) may be
replaced by the following ones:
\begin{equation}
P_{nlm_{l}}(r)\stackrel{r\to0}{\longrightarrow}0,
\qquad
P_{nlm_{l}}(r)\stackrel{r\to\infty}{\longrightarrow}0.
\label{2.8c}
\end{equation}
\label{2.8}%
\end{subequations}
The symbol $n$ that has appeared the first time as a subscript in
Eq.\ (\ref{2.6}) is the principal quantum number defined as
\begin{equation}
n=n_{r}+l+1,
\label{2.9}
\end{equation}
where $n_{r}\in\mathbb{N}_{0}$ is the radial quantum number which
counts the number of nodes (zeroes) in the radial wave function.

Since the term linear in $B$ which appears in the differential
operator in Eq.\ (\ref{2.8a}) is independent of the variable $r$, it
is clear that the energy eigenvalue $E_{nlm_{l}}$ may be written as
\begin{equation}
E_{nlm_{l}}=E_{nl}+E_{m_{l}}^{(1)},
\label{2.10}
\end{equation}
with
\begin{equation}
E_{m_{l}}^{(1)}=m_{l}\frac{e\hbar B}{2m}.
\label{2.11}
\end{equation}
It is also evident that the radial function $P_{nlm_{l}}(r)$ does
depend on $m_{l}$ through $l=|m_{l}|$ only:
\begin{equation}
P_{nlm_{l}}(r)\equiv P_{nl}(r).
\label{2.12}
\end{equation}
Consequently, the starting point for further considerations will be
the radial eigenvalue problem
\begin{subequations}
\begin{equation}
\left[-\frac{\hbar^{2}}{2m}\frac{\mathrm{d}^{2}}{\mathrm{d}r^{2}}
+\frac{\hbar^{2}\left(l^{2}-\frac{1}{4}\right)}{2mr^{2}}
-\frac{Ze^{2}}{(4\pi\epsilon_{0})r}
+\frac{e^{2}B^{2}r^{2}}{8m}\right]P_{nl}(r)=E_{nl}P_{nl}(r),
\label{2.13a}
\end{equation}
\begin{equation}
P_{nl}(r)\stackrel{r\to0}{\longrightarrow}0,
\qquad
P_{nl}(r)\stackrel{r\to\infty}{\longrightarrow}0.
\label{2.13b}
\end{equation}
\label{2.13}%
\end{subequations}
\section{Perturbation-theory analysis}
\label{III}
\setcounter{equation}{0}
\subsection{Basics and the zeroth-order problem}
\label{III.1}
Closed-form analytical solutions to the eigenproblem (\ref{2.13}) are
not known. Therefore, below we shall attempt to find its approximate
solutions, under the assumption that the magnetic field is weak, with
the use of the Rayleigh--Schr{\"o}dinger perturbation theory. To this
end, we write the radial differential operator from Eq.\
(\ref{2.13a}) as
\begin{equation}
H_{l}(r)=H_{l}^{(0)}(r)+H^{(2)}(r),
\label{3.1}
\end{equation}
where
\begin{equation}
H_{l}^{(0)}(r)
=-\frac{\hbar^{2}}{2m}\frac{\mathrm{d}^{2}}{\mathrm{d}r^{2}}
+\frac{\hbar^{2}\left(l^{2}-\frac{1}{4}\right)}{2mr^{2}}
-\frac{Ze^{2}}{(4\pi\epsilon_{0})r}
\label{3.2}
\end{equation}
and
\begin{equation}
H^{(2)}(r)=\frac{e^{2}B^{2}r^{2}}{8m}.
\label{3.3}
\end{equation}
We shall treat the diamagnetic term (\ref{3.3}) as a small
perturbation of the radial Coulomb Hamiltonian (\ref{3.2}). Since
$H^{(2)}(r)$ is of the \emph{second\/} order in the perturbing
magnetic field, we seek solutions to the eigensystem (\ref{2.13}) in
the form of the perturbation series
\begin{equation}
E_{nl}=E_{nl}^{(0)}
+E_{nl}^{(2)}+E_{nl}^{(4)}+\cdots
\label{3.4}
\end{equation}
and
\begin{equation}
P_{nl}(r)=P_{nl}^{(0)}(r)
+P_{nl}^{(2)}(r)+P_{nl}^{(4)}(r)+\cdots,
\label{3.5}
\end{equation}
involving \emph{even}-order terms only. Here $E_{nl}^{(0)}$ and
$P_{nl}^{(0)}(r)$ are those solutions to the zeroth-order
eigenproblem (being the radial Coulomb one)
\begin{subequations}
\begin{equation}
\big[H_{l}^{(0)}(r)-E^{(0)}\big]P^{(0)}(r)=0,
\label{3.6a}
\end{equation}
\begin{equation}
P^{(0)}(r)\stackrel{r\to0}{\longrightarrow}0,
\qquad
\textrm{$P^{(0)}(r)$ bounded for $r\to\infty$}
\label{3.6b}
\end{equation}
\label{3.6}%
\end{subequations}
(subscripts have been omitted intentionally), which correspond to the
\emph{discrete\/} part of its spectrum, consisting of the eigenvalues
\begin{equation}
E_{nl}^{(0)}\equiv E_{n}^{(0)}
=-\frac{Z^{2}}{2N_{n}^{2}}\frac{e^{2}}{(4\pi\epsilon_{0})a_{0}},
\label{3.7}
\end{equation}
with
\begin{equation}
N_{n}=n-{\textstyle\frac{1}{2}}
=n_{r}+l+{\textstyle\frac{1}{2}},
\label{3.8}
\end{equation}
and with
\begin{equation}
a_{0}=(4\pi\epsilon_{0})\frac{\hbar^{2}}{me^{2}}
\label{3.9}
\end{equation}
being the Bohr radius. Eigenfunctions associated with the eigenvalues
(\ref{3.7}), orthonormal in the sense of
\begin{equation}
\int_{0}^{\infty}\mathrm{d}r\:
P_{nl}^{(0)}(r)P_{n'l}^{(0)}(r)=\delta_{nn'},
\label{3.10}
\end{equation}
are
\begin{equation}
P_{nl}^{(0)}(r)
=\sqrt{\frac{Z(n-l-1)!}{a_{0}N_{n}^{2}(n+l-1)!}}
\left(\frac{2Zr}{N_{n}a_{0}}\right)^{l+1/2}
\mathrm{e}^{-Zr/N_{n}a_{0}}
L_{n-l-1}^{(2l)}\left(\frac{2Zr}{N_{n}a_{0}}\right),
\label{3.11}
\end{equation}
where $L_{k}^{(\alpha)}(x)$ is the generalized Laguerre polynomial
\cite[Sec.\ 5.5]{Magn66}. For integration purposes, it is frequently
convenient to have these functions rewritten as
\begin{equation}
P_{nl}^{(0)}(r)
=\sqrt{\frac{Zn_{r}!}{a_{0}N_{n}^{2}(n_{r}+2l)!}}
\left(\frac{2Zr}{N_{n}a_{0}}\right)^{l+1/2}
\mathrm{e}^{-Zr/N_{n}a_{0}}
L_{n_{r}}^{(2l)}\left(\frac{2Zr}{N_{n}a_{0}}\right).
\label{3.12}
\end{equation}
\subsection{The second-order corrections to Coulomb energies}
\label{III.2}
For the present problem, the second-order correction to energy,
$E_{nl}^{(2)}$, is given by
\begin{equation}
E_{nl}^{(2)}=\int_{0}^{\infty}\mathrm{d}r\:
P_{nl}^{(0)}(r)H^{(2)}(r)
P_{nl}^{(0)}(r),
\label{3.13}
\end{equation}
or equivalently, if use is made of Eq.\ (\ref{3.3}), by
\begin{equation}
E_{nl}^{(2)}=\frac{e^{2}B^{2}}{8m}
\int_{0}^{\infty}\mathrm{d}r\:r^{2}\big[P_{nl}^{(0)}(r)\big]^{2}.
\label{3.14}
\end{equation}
Plugging Eq.\ (\ref{3.12}) into the integrand and exploiting the
integration formula
\begin{eqnarray}
&& \int_{0}^{\infty}\mathrm{d}x\:x^{\alpha+3}\mathrm{e}^{-x}
\big[L_{k}^{(\alpha)}(x)\big]^{2}
=(2k+\alpha+1)(10k^{2}+10k+10\alpha k+\alpha^{2}+5\alpha+6)
\frac{\Gamma(k+\alpha+1)}{k!}
\nonumber \\
&& \hspace*{25em} (\Real\alpha>-4),
\label{3.15}
\end{eqnarray}
which may be deduced from the general expression \cite[Eqs.\ (E54),
(E56) and (E60)]{Szmy97}
\begin{eqnarray}
&& \int_{0}^{\infty}\mathrm{d}x\:x^{\gamma}\mathrm{e}^{-x}
L_{k}^{(\alpha)}(x)L_{k'}^{(\beta)}(x)
=(-)^{k+k'}\sum_{m=0}^{\min(k,k')}\frac{\Gamma(m+\gamma+1)}{m!}
\binom{\gamma-\alpha}{k-m}\binom{\gamma-\beta}{k'-m}
\nonumber \\
&& \hspace*{25em} (\Real\gamma>-1),
\label{3.16}
\end{eqnarray}
yields
\begin{equation}
E_{nl}^{(2)}=\frac{1}{2^{4}}
\left(n-{\textstyle\frac{1}{2}}\right)^{2}
\left(5n^{2}-5n-3l^{2}+3\right)Z^{-2}
\frac{B^{2}}{B_{0}^{2}}\frac{e^{2}}{(4\pi\epsilon_{0})a_{0}},
\label{3.17}
\end{equation}
where
\begin{equation}
B_{0}=\frac{\hbar}{ea_{0}^{2}}
=\frac{m^{2}e^{3}}{(4\pi\epsilon_{0})^{2}\hbar^{3}}
\label{3.18}
\end{equation}
is the atomic unit of magnetic induction. For states with $l=n-1$
(i.e., those with $n_{r}=0$), the expression in Eq.\ (\ref{3.17})
simplifies to
\begin{equation}
E_{n,n-1}^{(2)}=\frac{1}{2^{3}}n
\left(n+{\textstyle\frac{1}{2}}\right)
\left(n-{\textstyle\frac{1}{2}}\right)^{2}Z^{-2}
\frac{B^{2}}{B_{0}^{2}}\frac{e^{2}}{(4\pi\epsilon_{0})a_{0}}.
\label{3.19}
\end{equation}
\subsection{The fourth-order corrections to Coulomb energies} 
\label{III.3}
Proceeding along the standard route, one finds that for the present
problem the fourth-order correction to energy, $E_{nl}^{(4)}$, is
given by
\begin{equation}
E_{nl}^{(4)}=\int_{0}^{\infty}\mathrm{d}r\:
P_{nl}^{(0)}(r)H^{(2)}(r)P_{nl}^{(2)}(r),
\label{3.20}
\end{equation}
where the second-order correction to the radial wave function,
$P_{nl}^{(2)}(r)$, is a solution to the inhomogeneous boundary-value
problem
\begin{subequations}
\begin{equation}
\big[H_{l}^{(0)}(r)-E_{n}^{(0)}\big]P_{nl}^{(2)}(r)
=-\big[H^{(2)}(r)-E_{nl}^{(2)}\big]P_{nl}^{(0)}(r)
\label{3.21a}
\end{equation}
\begin{equation}
P_{nl}^{(2)}(r)\stackrel{r\to0}{\longrightarrow}0,
\qquad
P_{nl}^{(2)}(r)\stackrel{r\to\infty}{\longrightarrow}0,
\label{3.21b}
\end{equation}
\label{3.21}%
\end{subequations}
subject to the further orthogonality restraint
\begin{equation}
\int_{0}^{\infty}\mathrm{d}r\:
P_{nl}^{(0)}(r)P_{nl}^{(2)}(r)=0.
\label{3.22}
\end{equation}
The formal solution to the problem (\ref{3.21})--(\ref{3.22}) is
\begin{equation}
P_{nl}^{(2)}(r)=-\int_{0}^{\infty}\mathrm{d}r'\:
\widetilde{G}_{nl}^{(0)}(r,r')\big[H^{(2)}(r')-E_{nl}^{(2)}\big]
P_{nl}^{(0)}(r'),
\label{3.23}
\end{equation}
where $\widetilde{G}_{nl}^{(0)}(r,r')$ is a generalized (or reduced)
radial Coulomb Green function associated with the Coulomb energy
level $E_{n}^{(0)}$. The latter function is defined as that
particular solution to the inhomogeneous boundary-value problem
\begin{subequations}
\begin{equation}
\big[H_{l}^{(0)}(r)-E_{n}^{(0)}\big]\widetilde{G}_{nl}^{(0)}(r,r')
=\delta(r-r')-P_{nl}^{(0)}(r)P_{nl}^{(0)}(r'),
\label{3.24a}
\end{equation}
\begin{equation}
\widetilde{G}_{nl}^{(0)}(r,r')\stackrel{r\to0}{\longrightarrow}0,
\qquad
\widetilde{G}_{nl}^{(0)}(r,r')\stackrel{r\to\infty}{\longrightarrow}0,
\label{3.24b}
\end{equation}
\label{3.24}%
\end{subequations}
where $\delta(r-r')$ is the Dirac delta function, which obeys the
additional orthogonality constraint
\begin{equation}
\int_{0}^{\infty}\mathrm{d}r\:
P_{nl}^{(0)}(r)\widetilde{G}_{nl}^{(0)}(r,r')=0.
\label{3.25}
\end{equation}
Since the zeroth-order eigenproblem (\ref{3.6}) is self-adjoint, the
function $\widetilde{G}_{nl}^{(0)}(r,r')$ is symmetric in its
arguments:
\begin{equation}
\widetilde{G}_{nl}^{(0)}(r,r')=\widetilde{G}_{nl}^{(0)}(r',r).
\label{3.26}
\end{equation}
When this is combined with Eq.\ (\ref{3.25}), one deduces the formula
\begin{equation}
\int_{0}^{\infty}\mathrm{d}r'\:
\widetilde{G}_{nl}^{(0)}(r,r')P_{nl}^{(0)}(r')=0,
\label{3.27}
\end{equation}
which allows us to simplify Eq.\ (\ref{3.23}) to obtain
\begin{equation}
P_{nl}^{(2)}(r)=-\int_{0}^{\infty}\mathrm{d}r'\:
\widetilde{G}_{nl}^{(0)}(r,r')H^{(2)}(r')P_{nl}^{(0)}(r').
\label{3.28}
\end{equation}
Plugging Eq.\ (\ref{3.28}) into the right-hand side of Eq.\
(\ref{3.20}) gives the energy correction $E_{nl}^{(4)}$ in the form
\begin{equation}
E_{nl}^{(4)}=-\int_{0}^{\infty}\mathrm{d}r
\int_{0}^{\infty}\mathrm{d}r'\:P_{nl}^{(0)}(r)
H^{(2)}(r)\widetilde{G}_{nl}^{(0)}(r,r')
H^{(2)}(r')P_{nl}^{(0)}(r')
\label{3.29}
\end{equation}
or, still more explicitly, in the form
\begin{equation}
E_{nl}^{(4)}=-\left(\frac{e^{2}B^{2}}{8m}\right)^{2}
\int_{0}^{\infty}\mathrm{d}r\int_{0}^{\infty}\mathrm{d}r'\:
P_{nl}^{(0)}(r)r^{2}\widetilde{G}_{nl}^{(0)}(r,r')r^{\prime\,2}
P_{nl}^{(0)}(r').
\label{3.30}
\end{equation}

A representation of the generalized radial Coulomb Green function
$\widetilde{G}_{nl}^{(0)}(r,r')$ which is perhaps the most suitable
for the use in Eq.\ (\ref{3.30}) is the one in the form of a series
expansion in the discrete radial Coulomb Sturmian basis. We shall
construct it below.

The discrete radial Coulomb Sturmian functions are defined as
solutions to the spectral problem
\begin{subequations}
\begin{equation}
\left[-\frac{\hbar^{2}}{2m}\frac{\mathrm{d}^{2}}{\mathrm{d}r^{2}}
+\frac{\hbar^{2}(l^{2}-\frac{1}{4})}{2mr^{2}}
-\mu_{n_{r}l}^{(0)}(E)\frac{Ze^{2}}{(4\pi\epsilon_{0})r}-E\right]
S_{n_{r}l}^{(0)}(E,r)=0
\qquad (E<0),
\label{3.31a}
\end{equation}
\begin{equation}
S_{n_{r}l}^{(0)}(E,r)\stackrel{r\to0}{\longrightarrow}0,
\qquad
S_{n_{r}l}^{(0)}(E,r)\stackrel{r\to\infty}{\longrightarrow}0,
\label{3.31b}
\end{equation}
\label{3.31}%
\end{subequations}
with $E<0$ fixed and with the parameter $\mu_{n_{r}l}^{(0)}(E)$
chosen as an eigenvalue. The spectrum of this problem is purely
discrete, and eigenvalues are given by
\begin{equation}
\mu_{n_{r}l}^{(0)}(E)
=\left(n_{r}+l+{\textstyle\frac{1}{2}}\right)\frac{ka_{0}}{Z}
\qquad (n_{r}\in\mathbb{N}_{0}),
\label{3.32}
\end{equation}
where
\begin{equation}
k=\sqrt{-\frac{2mE}{\hbar^{2}}}.
\label{3.33}
\end{equation}
Eigenfunctions, orthonormal in the sense of
\begin{equation}
\int_{0}^{\infty}\mathrm{d}r\:\frac{Ze^{2}}{(4\pi\epsilon_{0})r}
S_{n_{r}l}^{(0)}(E,r)S_{n_{r}^{\prime}l}^{(0)}(E,r)
=\delta_{n_{r}n_{r}^{\prime}},
\label{3.34}
\end{equation}
are
\begin{equation}
S_{n_{r}l}^{(0)}(E,r)
=\sqrt{\frac{(4\pi\epsilon_{0})n_{r}!}{Ze^{2}(n_{r}+2l)!}}\,
(2kr)^{l+1/2}\mathrm{e}^{-kr}L_{n_{r}}^{(2l)}(2kr).
\label{3.35}
\end{equation}
In contrary to the discrete Coulomb eigenfunctions (\ref{3.11}), the
Sturmians (\ref{3.35}) form a complete set, the corresponding closure
relation being
\begin{equation}
\frac{Ze^{2}}{(4\pi\epsilon_{0})r}
\sum_{n_{r}=0}^{\infty}S_{n_{r}l}^{(0)}(E,r)S_{n_{r}l}^{(0)}(E,r')
=\delta(r-r').
\label{3.36}
\end{equation}
If the parameter $E$ coincides with the Coulomb energy eigenvalue
$E_{n}^{(0)}$ displayed in Eq.\ (\ref{3.7}) [we assume $n$ is related
to $n_{r}$ and $l$ used here as in Eq.\ (\ref{2.9})], it is easy to
see from Eqs.\ (\ref{3.32}), (\ref{3.33}), (\ref{3.7}) and
(\ref{3.8}) that one has
\begin{equation}
\mu_{n_{r}l}^{(0)}(E_{n}^{(0)})=1
\qquad (n=n_{r}+l+1).
\label{3.37}
\end{equation}
Similarly, from Eqs.\ (\ref{3.35}), (\ref{3.31}), (\ref{3.7}),
(\ref{3.8}) and (\ref{3.12}) one infers the relationship
\begin{equation}
S_{n_{r}l}^{(0)}(E_{n}^{(0)},r)=\frac{N_{n}}{Z}
\sqrt{\frac{(4\pi\epsilon_{0})a_{0}}{e^{2}}}P_{nl}^{(0)}(r)
\qquad (n=n_{r}+l+1).
\label{3.38}
\end{equation}

The radial Coulomb Green function, $G_{l}^{(0)}(E,r,r')$, is defined
to be a solution to the inhomogeneous equation
\begin{subequations}
\begin{equation}
\big[H_{l}^{(0)}(r)-E\big]G_{l}^{(0)}(E,r,r')=\delta(r-r')
\qquad (E<0),
\label{3.39a}
\end{equation}
subject to the boundary constraints
\begin{equation}
G_{l}^{(0)}(E,r,r')\stackrel{r\to0}{\longrightarrow}0,
\qquad
G_{l}^{(0)}(E,r,r')\stackrel{r\to\infty}{\longrightarrow}0.
\label{3.39b}
\end{equation}
\label{3.39}%
\end{subequations}
Since the Sturmian functions (\ref{3.35}) form a complete set, the
Green function $G_{l}^{(0)}(E,r,r')$ may be sought in the form of the
series
\begin{equation}
G_{l}^{(0)}(E,r,r')=\sum_{n_{r}=0}^{\infty}
C_{n_{r}l}^{(0)}(E,r')S_{n_{r}l}^{(0)}(E,r).
\label{3.40}
\end{equation}
To determine the expansion coefficients $C_{n_{r}l}^{(0)}(E,r)$, we
plug Eq.\ (\ref{3.40}) into Eq.\ (\ref{3.39a}), multiply both sides
of the resulting identity with $S_{n_{r}^{\prime}l}^{(0)}(E,r)$, then
integrate with respect to $r$ over the interval $[0,\infty)$, and
apply the orthogonality relation (\ref{3.35}). Upon the replacement
of $n_{r}^{\prime}$ with $n_{r}$, this yields
\begin{equation}
C_{n_{r}l}^{(0)}(E,r')=\frac{1}{\mu_{n_{r}l}^{(0)}(E)-1}
S_{n_{r}l}^{(0)}(E,r'),
\label{3.41}
\end{equation}
hence, we obtain the following symmetric Sturmian expansion of
$G_{l}^{(0)}(E,r,r')$:
\begin{equation}
G_{l}^{(0)}(E,r,r')=\sum_{n_{r}=0}^{\infty}
\frac{S_{n_{r}l}^{(0)}(E,r)S_{n_{r}l}^{(0)}(E,r')}
{\mu_{n_{r}l}^{(0)}(E)-1}.
\label{3.42}
\end{equation}

It follows from Eqs.\ (\ref{3.24}), (\ref{3.25}) and (\ref{3.39})
that the generalized radial Coulomb Green function
$\widetilde{G}_{nl}^{(0)}(r,r')$ may be obtained from the radial
Coulomb Green function $G_{l}^{(0)}(E,r,r')$ through the limit
procedure
\begin{equation}
\widetilde{G}_{nl}^{(0)}(r,r')
=\lim_{E\to E_{n}^{(0)}}\left[G_{l}^{(0)}(E,r,r')
-\frac{P_{nl}^{(0)}(r)P_{nl}^{(0)}(r')}{E_{n}^{(0)}-E}\right].
\label{3.43}
\end{equation}
By virtue of the de l'Hospital rule, the latter equation is
equivalent to the following one:
\begin{equation}
\widetilde{G}_{nl}^{(0)}(r,r')=\lim_{E\to E_{n}^{(0)}}
\frac{\partial}{\partial E}
\big[\big(E-E_{n}^{(0)}\big)G_{l}^{(0)}(E,r,r')\big],
\label{3.44}
\end{equation}
which is particularly suitable for the construction of the Sturmian
expansion of $\widetilde{G}_{nl}^{(0)}(r,r')$. Inserting the series
representation (\ref{3.42}) into the right-hand side of Eq.\
(\ref{3.44}) and then making use of the relationships
\begin{equation}
\frac{\partial S_{n_{r}l}^{(0)}(E,r)}{\partial E}
=\frac{r}{2E}\frac{\mathrm{d}S_{n_{r}l}^{(0)}(E,r)}{\mathrm{d}r},
\label{3.45}
\end{equation}
\begin{equation}
\frac{E-E_{n}^{(0)}}{\mu_{n_{r}l}^{(0)}(E)-1}
=E_{n}^{(0)}\big[\mu_{n_{r}l}^{(0)}(E)+1\big],
\label{3.46}
\end{equation}
\begin{equation}
\lim_{E\to E_{n}^{(0)}}
\frac{E-E_{n}^{(0)}}{\mu_{n_{r}l}^{(0)}(E)-1}=2E_{n}^{(0)}
\qquad (n=n_{r}+l+1),
\label{3.47}
\end{equation}
\begin{equation}
\lim_{E\to E_{n}^{(0)}}\frac{\partial}{\partial E}
\frac{E-E_{n}^{(0)}}{\mu_{n_{r}l}^{(0)}(E)-1}=\frac{1}{2}
\qquad (n=n_{r}+l+1),
\label{3.48}
\end{equation}
\begin{equation}
\mu_{n_{r}^{\prime}l}^{(0)}(E_{n}^{(0)})
=\frac{n_{r}^{\prime}+l+\frac{1}{2}}{N_{n}},
\label{3.49}
\end{equation}
which may be easily derived from the defining Eqs.\ (\ref{3.32}) and
(\ref{3.35}), one eventually arrives at the sought Sturmian expansion
of the generalized radial Coulomb Green function, which is
\begin{eqnarray}
\widetilde{G}_{nl}^{(0)}(r,r') 
&=& N_{n}\sum_{\substack{n_{r}^{\prime}=0 \\ 
(n_{r}^{\prime}\neq n_{r})}}^{\infty}
\frac{S_{n_{r}^{\prime}l}^{(0)}(E_{n}^{(0)},r)
S_{n_{r}^{\prime}l}^{(0)}(E_{n}^{(0)},r')}
{n_{r}^{\prime}-n_{r}}
+\frac{1}{2}S_{n_{r}l}^{(0)}(E_{n}^{(0)},r)
S_{n_{r}l}^{(0)}(E_{n}^{(0)},r')
\nonumber \\
&& +\,r\frac{\mathrm{d}S_{n_{r}l}^{(0)}(E_{n}^{(0)},r)}
{\mathrm{d}r}S_{n_{r}l}^{(0)}(E_{n}^{(0)},r')
+S_{n_{r}l}^{(0)}(E_{n}^{(0)},r)
r'\frac{\mathrm{d}S_{n_{r}l}^{(0)}(E_{n}^{(0)},r')}{\mathrm{d}r'}
\nonumber \\
&& \hspace*{20em} (n_{r}=n-l-1).
\label{3.50}
\end{eqnarray}

Once the Sturmian expansion of $\widetilde{G}_{nl}^{(0)}(r,r')$ has
been found, we are ready to complete the task to find the
fourth-order energy correction $E_{nl}^{(4)}$. To this end, we insert
Eq.\ (\ref{3.50}) into Eq.\ (\ref{3.30}) and use the relationship in
Eq.\ (\ref{3.38}), together with integrations by parts, to eliminate
derivatives of Sturmian functions. This gives $E_{nl}^{(4)}$ in the
form
\begin{eqnarray}
E_{nl}^{(4)} &=& -\left(\frac{e^{2}B^{2}}{8m}\right)^{2}
\Bigg\{N_{n}\sum_{\substack{n_{r}^{\prime}=0 \\ 
(n_{r}^{\prime}\neq n_{r})}}^{\infty}
\frac{\displaystyle\left[\int_{0}^{\infty}\mathrm{d}r\:
r^{2}P_{nl}^{(0)}(r)
S_{n_{r}^{\prime}l}^{(0)}(E_{n}^{(0)},r)\right]^{2}}
{n_{r}^{\prime}-n_{r}}
\nonumber \\
&& -\,\frac{5}{2}\bigg[\int_{0}^{\infty}\mathrm{d}r\:r^{2}
P_{nl}^{(0)}(r)S_{n_{r}l}^{(0)}(E_{n}^{(0)},r)\bigg]^{2}\Bigg\}
\qquad (n_{r}=n-l-1).
\label{3.51}
\end{eqnarray}
The integrals in Eq.\ (\ref{3.51}) may be taken after one exploits
Eqs.\ (\ref{3.12}) and (\ref{3.35}), with the use of the integration
formula
\begin{eqnarray}
&& \hspace*{-5em}
\int_{0}^{\infty}\mathrm{d}x\:x^{\alpha+3}\mathrm{e}^{-x}
L_{k}^{(\alpha)}(x)L_{k'}^{(\alpha)}(x)
\nonumber \\
&=& -\,\frac{\Gamma(k+\alpha+1)}{(k-3)!}\delta_{k',k-3}
+3(2k+\alpha-1)\frac{\Gamma(k+\alpha+1)}{(k-2)!}\delta_{k',k-2}
\nonumber \\
&& -\,3(5k^2+5\alpha k+\alpha^{2}+1)\frac{\Gamma(k+\alpha+1)}{(k-1)!}
\delta_{k',k-1}
\nonumber \\
&& +\,(2k+\alpha+1)(10k^{2}+10k+10\alpha k+\alpha^{2}+5\alpha+6)
\frac{\Gamma(k+\alpha+1)}{k!}\delta_{k'k}
\nonumber \\
&& -\,3(5k^{2}+10k+5\alpha k+\alpha^{2}+5\alpha+6)
\frac{\Gamma(k+\alpha+2)}{k!}\delta_{k',k+1}
\nonumber \\
&& +\,3(2k+\alpha+3)\frac{\Gamma(k+\alpha+3)}{k!}\delta_{k',k+2}
-\frac{\Gamma(k+\alpha+4)}{k!}\delta_{k',k+3}
\qquad (\Real\alpha>-4),
\label{3.52}
\end{eqnarray}
which generalizes the one in Eq.\ (\ref{3.15}) and, similarly to the
latter, may be derived from the general expression (\ref{3.16}).
Since only terms with $n_{r}^{\prime}$ constrained by
$1\leqslant|n_{r}^{\prime}-n_{r}|\leqslant3$ are seen to contribute
non-vanishingly to the sum in Eq.\ (\ref{3.51}), we eventually obtain
\begin{eqnarray}
E_{nl}^{(4)} &=& -\frac{1}{2^{10}}
\left(n-{\textstyle\frac{1}{2}}\right)^{6}
\nonumber \\
&& \times\left(143n^{4}-286n^{3}-90n^{2}l^{2}+582n^{2}+90nl^{2}-439n
-21l^{4}-138l^{2}+159\right)
\nonumber \\
&& \times\,Z^{-6}\frac{B^{4}}{B_{0}^{4}}
\frac{e^{2}}{(4\pi\epsilon_{0})a_{0}}.
\label{3.53}
\end{eqnarray}
For states with $l=n-1$ (i.e., those with $n_{r}=0$), Eq.\
(\ref{3.53}) becomes
\begin{equation}
E_{n,n-1}^{(4)}=-\frac{1}{2^{9}}n
\left(n+{\textstyle\frac{1}{2}}\right)
\left(n-{\textstyle\frac{1}{2}}\right)^{6}(16n^{2}+26n+11)
Z^{-6}\frac{B^{4}}{B_{0}^{4}}\frac{e^{2}}{(4\pi\epsilon_{0})a_{0}}.
\label{3.54}
\end{equation}
For the ground state ($n=1$), Eq.\ (\ref{3.54}) yields
\begin{equation}
E_{10}^{(4)}=-\frac{159}{65\,536}Z^{-6}\frac{B^{4}}{B_{0}^{4}}
\frac{e^{2}}{(4\pi\epsilon_{0})a_{0}}.
\label{3.55}
\end{equation}
This differs from the result announced in Refs.\ \cite[Eq.\
(32)]{Hoan13a} and \cite[Eq.\ (6.59)]{Fera15}, which is
\begin{equation}
E_{10}^{(4)}=-\frac{153}{65\,536}Z^{-6}\frac{B^{4}}{B_{0}^{4}}
\frac{e^{2}}{(4\pi\epsilon_{0})a_{0}}.
\label{3.56}
\end{equation}
The latter one is thus found to be incorrect.
%
%
\section{Summary and concluding remarks}
\label{IV}
\setcounter{equation}{0}
On the preceding pages, we have shown that energy levels of the
planar hydrogen-like atom placed in a weak, static, uniform magnetic
field of induction $\boldsymbol{B}$ perpendicular to the atomic plane
may be expressed in the form
\begin{equation}
E_{nlm_{l}}=E_{n}^{(0)}+E_{m_{l}}^{(1)}
+E_{nl}^{(2)}+E_{nl}^{(4)}+O\left(Z^{-10}(B/B_{0})^{6}\right),
\label{4.1}
\end{equation}
where
\begin{equation}
E_{\ldots}^{(k)}=\varepsilon_{\ldots}^{(k)}Z^{-2k+2}
\frac{B^{k}}{B_{0}^{k}}\frac{e^{2}}{(4\pi\epsilon_{0})a_{0}}.
\label{4.2}
\end{equation}
In Eq.\ (\ref{4.2}), $Z$ is an electric charge of the atomic nucleus
in units of the elementary charge $e$, $a_{0}$ is the Bohr radius,
\begin{equation}
B_{0}=\frac{m^{2}e^{3}}{(4\pi\epsilon_{0})^{2}\hbar^{3}}
\simeq2.35\times10^{5}~\textrm{T}
\label{4.3}
\end{equation}
is the atomic unit of magnetic induction, while the dimensionless and
$Z$-independent coefficients $\varepsilon_{\ldots}^{(k)}$ are given
by
\begin{equation}
\varepsilon_{n}^{(0)}=-\frac{1}{2\left(n-\frac{1}{2}\right)^{2}},
\label{4.4}
\end{equation}
\begin{equation}
\varepsilon_{m_{l}}^{(1)}=\frac{1}{2}m_{l},
\label{4.5}
\end{equation}
\begin{equation}
\varepsilon_{nl}^{(2)}=\frac{1}{2^{4}}
\left(n-{\textstyle\frac{1}{2}}\right)^{2}
\left(5n^{2}-5n-3l^{2}+3\right),
\label{4.6}
\end{equation}
and
\begin{eqnarray}
&& \hspace*{-2em}
\varepsilon_{nl}^{(4)}=-\frac{1}{2^{10}}
\left(n-{\textstyle\frac{1}{2}}\right)^{6}
\left(143n^{4}-286n^{3}-90n^{2}l^{2}+582n^{2}+90nl^{2}-439n
-21l^{4}-138l^{2}+159\right),
\nonumber \\
&&
\label{4.7}
\end{eqnarray}
with $n\in\mathbb{N}_{+}$, $m_{l}\in\mathbb{Z}$ and $0\leqslant
l=|m_{l}|\leqslant n-1$. Numerical values of the coefficients
$\varepsilon_{nl}^{(2)}$ and $\varepsilon_{nl}^{(4)}$ for states with
$1\leqslant n\leqslant4$ are displayed in Table \ref{T.1}.
\begin{table}[h]
\caption{Numerical values of the coefficients
$\varepsilon_{nl}^{(2)}$ and $\varepsilon_{nl}^{(4)}$, defined in
Eqs.\ (\ref{4.6}) and (\ref{4.7}), for $1\leqslant n\leqslant4$ and
$0\leqslant l\leqslant n-1$.}
\label{T.1}
\begin{center}
\begin{tabular}{cccccccccc}
\hline \\*[-1ex]
$n$ & $l$ 
&& \multicolumn{3}{c}{$\varepsilon_{nl}^{(2)}$}
&& \multicolumn{3}{c}{$\varepsilon_{nl}^{(4)}$} \\*[0.5ex]
\cline{4-6} \cline{8-10} \\*[-1.25ex]
&&& rational form && factorized form
&& rational form && factorized form \\*[1ex]
\hline \\*[-1ex]
1 & 0 
&& $\displaystyle\frac{3}{64}$
&& $\displaystyle\frac{3}{2^{6}}$
&& $\displaystyle-\frac{159}{65\,536}$ 
&& $\displaystyle-\frac{3\times53}{2^{16}}$ 
\\*[2ex]
2 & 0
&& $\displaystyle\frac{117}{64}$
&& $\displaystyle\frac{3^{2}\times13}{2^{6}}$
&& $\displaystyle-\frac{1\,172\,961}{65\,536}$ 
&& $\displaystyle-\frac{3^{6}\times1609}{2^{16}}$ 
\\*[2ex]
  & 1
&& $\displaystyle\frac{45}{32}$
&& $\displaystyle\frac{3^{2}\times5}{2^{5}}$
&& $\displaystyle-\frac{462\,915}{32\,768}$ 
&& $\displaystyle-\frac{3^{6}\times5\times127}{2^{15}}$ 
\\*[2ex]
3 & 0
&& $\displaystyle\frac{825}{64}$
&& $\displaystyle\frac{3\times5^{2}\times11}{2^{6}}$
&& $\displaystyle-\frac{124\,078\,125}{65\,536}$
&& $\displaystyle-\frac{3\times5^{6}\times2647}{2^{16}}$
\\*[2ex]
  & 1
&& $\displaystyle\frac{375}{32}$
&& $\displaystyle\frac{3\times5^{3}}{2^{5}}$
&& $\displaystyle-\frac{56\,578\,125}{32\,768}$ 
&& $\displaystyle-\frac{3\times5^{6}\times17\times71}{2^{15}}$
\\*[2ex]
  & 2
&& $\displaystyle\frac{525}{64}$
&& $\displaystyle\frac{3\times5^{2}\times7}{2^{6}}$
&& $\displaystyle-\frac{76\,453\,125}{65\,536}$ 
&& $\displaystyle-\frac{3\times5^{6}\times7\times233}{2^{16}}$ 
\\*[2ex]
4 & 0
&& $\displaystyle\frac{3087}{64}$
&& $\displaystyle\frac{3^{2}\times7^{3}}{2^{6}}$
&& $\displaystyle-\frac{3\,061\,109\,331}{65\,536}$ 
&& $\displaystyle-\frac{3^{2}\times7^{8}\times59}{2^{16}}$ 
\\*[2ex]
  & 1
&& $\displaystyle\frac{735}{16}$
&& $\displaystyle\frac{3\times5\times7^{2}}{2^{4}}$
&& $\displaystyle-\frac{728\,835\,555}{16\,384}$ 
&& $\displaystyle-\frac{3\times5\times7^{7}\times59}{2^{14}}$ 
\\*[2ex]
  & 2
&& $\displaystyle\frac{2499}{64}$
&& $\displaystyle\frac{3\times7^{2}\times17}{2^{6}}$
&& $\displaystyle-\frac{2\,448\,393\,339}{65\,536}$ 
&& $\displaystyle-\frac{3\times7^{7}\times991}{2^{16}}$ 
\\*[2ex]
  & 3
&& $\displaystyle\frac{441}{16}$
&& $\displaystyle\frac{3^{2}\times7^{2}}{2^{4}}$
&& $\displaystyle-\frac{392\,830\,011}{16\,384}$ 
&& $\displaystyle-\frac{3^{2}\times7^{7}\times53}{2^{14}}$ 
\\*[2ex]
\hline
\end{tabular}
\end{center}
\end{table}

It has to be emphasized that the formula in Eq.\ (\ref{4.1}) is valid
only if the electron spin is ignored. If this cannot be done, the
Schr{\"o}dinger equation (\ref{2.1}) should be replaced with the
planar Pauli equation
\begin{equation}
\left\{\frac{\{\boldsymbol{\sigma}
\cdot[-\mathrm{i}\hbar\boldsymbol{\nabla}
+e\boldsymbol{A}(\boldsymbol{r})\}^{2}}{2m}
-\frac{Ze^{2}}{(4\pi\epsilon_{0})r}\right\}\Psi(\boldsymbol{r})
=E\Psi(\boldsymbol{r})
\qquad (\boldsymbol{r}\in\mathbb{R}^{2}),
\label{4.8}
\end{equation}
where $\boldsymbol{\sigma}=(\sigma_{x},\sigma_{y})$ is the
two-dimensional Pauli matrix vector, and $\Psi(\boldsymbol{r})$ is a
two-component Pauli spinor. Equation (\ref{4.8}) may be cast into the
form
\begin{equation}
\left\{\frac{[-\mathrm{i}\hbar\boldsymbol{\nabla}
+e\boldsymbol{A}(\boldsymbol{r})]^{2}}{2m}
+\frac{e\hbar B}{m}\Sigma_{z}
-\frac{Ze^{2}}{(4\pi\epsilon_{0})r}\right\}\Psi(\boldsymbol{r})
=E\Psi(\boldsymbol{r}),
\label{4.9}
\end{equation}
with
\begin{equation}
\Sigma_{z}=\frac{1}{2}\sigma_{z},
\label{4.10}
\end{equation}
where $\sigma_{z}$ is the third Pauli matrix. It is then evident that
Eq.\ (\ref{4.9}), supplemented with the regularity constraints on
$\Psi(\boldsymbol{r})$ analogous to those introduced under Eq.\
(\ref{2.1}), possesses separated eigenfunctions of the form
\begin{equation}
\Psi_{nlm_{l}m_{s}}(r,\varphi)
=\frac{1}{\sqrt{r}}P_{nl}(r)
\frac{\mathrm{e}^{\mathrm{i}m_{l}\varphi}}{\sqrt{2\pi}}\chi_{m_{s}},
\label{4.11}
\end{equation}
where $P_{nl}(r)$ is the same radial function which has appeared in
the preceding sections, while $\chi_{m_{s}}$ is the spin one-half
eigenfunction obeying
\begin{equation}
\Sigma_{z}\chi_{m_{s}}=m_{s}\chi_{m_{s}}
\qquad \left(m_{s}=\pm{\textstyle\frac{1}{2}}\right),
\label{4.12}
\end{equation}
and that the energy spectrum is of the form
\begin{equation}
E_{nlm_{l}m_{s}}=E_{n}^{(0)}+E_{m_{l}m_{s}}^{(1)}+E_{nl}^{(2)}
+E_{nl}^{(4)}+O\left(Z^{-10}(B/B_{0})^{6}\right),
\label{4.13}
\end{equation}
with the terms $E_{n}^{(0)}$, $E_{nl}^{(2)}$ and $E_{nl}^{(4)}$ being
identical to those derived before, and with
\begin{equation}
E_{m_{l}m_{s}}^{(1)}=\frac{1}{2}(m_{l}+2m_{s})
\frac{B}{B_{0}}\frac{e^{2}}{(4\pi\epsilon_{0})a_{0}}
\qquad \left(=(m_{l}+2m_{s})\frac{e\hbar B}{2m}\right).
\label{4.14}
\end{equation}
%
%

%
\end{document}